\documentclass[prd,twocolumn,nofootinbib,notitlepage,10pt,aps]{revtex4-1}
\usepackage[utf8]{inputenc}
\usepackage{cancel}
\usepackage{xcolor}
\usepackage{latexsym}
\usepackage{amsmath}
\usepackage{amssymb}
\usepackage{bbm}

\usepackage{hyperref}
\hypersetup{
    colorlinks,
    linkcolor={violet},
    citecolor={teal},
    urlcolor={teal}
}
\usepackage{orcidlink}

\usepackage{ulem}
\usepackage{pdfsync}
\usepackage{epsfig}
\usepackage{epstopdf}
\usepackage{subfigure}
\usepackage{color}
\usepackage{comment}
\usepackage{slashed}
\usepackage{physics}

\def\beq{\begin{equation}}
\def\eeq{\end{equation}}
\def\baq{\begin{eqnarray}}
\def\eaq{\end{eqnarray}}

\newcommand{\be}{\begin{equation}} % only untightened
\newcommand{\ee}{\end{equation}}
\newcommand{\bea}{\begin{eqnarray}} % only untightened
\newcommand{\eea}{\end{eqnarray}}

\newcommand{\bmp}{\noindent\begin{minipage}{16cm}}
\newcommand{\emp}{\end{minipage}\vskip 7mm} % 7mm untightened
\def\lsim{\mathrel{\raise.3ex\hbox{$<$\kern-.75em\lower1ex\hbox{$\sim$}}}}
\def\gsim{\mathrel{\raise.3ex\hbox{$>$\kern-.75em\lower1ex\hbox{$\sim$}}}}

\newcommand{\intron}[1]{}%{#1}

\begin{document}

%\title{Palatini Linear Attractors in Light of ACT}
%\title{Palatini in ACTion: Probing Linear Attractors in Cosmic Inflation}
%\title{An ACTive Role for Linear Attractors in Palatini Gravity}
\title{Palatini Linear Attractors Are Back in ACTion}
%\title{Linear Attractors in Palatini Inflation: An ACTive Probe of Early Universe Dynamics}
%\title{Palatini Linear ACTractors}

\author{Christian Dioguardi\orcidlink{0000-0002-6133-0383}}
\email{christian.dioguardi@kbfi.ee}
\affiliation{National Institute of Chemical Physics and Biophysics, \\
R\"avala 10, 10143 Tallinn, Estonia}
\affiliation{Tallinn University of Technology, Akadeemia tee 23, 12618 Tallinn, Estonia}

\author{Alexandros Karam\orcidlink{0000-0002-0582-8996}}
\email{alexandros.karam@kbfi.ee}
\affiliation{National Institute of Chemical Physics and Biophysics, \\
R\"avala 10, 10143 Tallinn, Estonia}

\begin{abstract}

Recent results from the Atacama Cosmology Telescope (ACT) indicate a scalar spectral index $n_s \simeq 0.9743$, in excellent agreement with the prediction of linear inflation. However, the corresponding tensor-to-scalar ratio $r \simeq 0.0667$ is in tension with current observational bounds. In this work, we investigate how this tension can be alleviated in the Palatini formulation of gravity. We consider two classes of models based on simple monomial potentials: (i) models with a non-minimal coupling between the inflaton and gravity, and (ii) models including an $\alpha R^2$ term. In the first case, we find that a quadratic potential with a linear non-minimal coupling leads to the linear inflation attractor, with $r$ suppressed as $\xi$ increases. In the second case, we show that a linear potential can yield values of $r$ consistent with observations for sufficiently large $\alpha$. Our results demonstrate that simple monomial models can remain compatible with current observational constraints when embedded in the Palatini framework.

\end{abstract}

%%%%%%%%%%%%%%%%%%%%%%%%%%%%%%%%%%%%%%%%%%%%%%%%%%%%%%%%%%%%%%%%%%%%%%%%%%%%%%%%%%%%%%%%%%%%%%%%%%%%
% DOCUMENT
%
\maketitle
%%%%%%%%%%%%%%%%%%%%%%%%%%%%%%%%%%%%%%%%%%%%%%%%%%%%%%%%%%%%%%%%%%%%%%%%%%%%%%%%%%%%%%%%%%%%%%%%%%%%

%%%%%%%%%%%%%%%%%%%%%%%%%%%%%%%%%%%%%%%%%%%%%%%%%%%%%%%%%%%%%%%%%%%%%%%%%%%%%%%%%%%%%%%%%%%%%%%%%%%%
%%%%%%%%%%%%%%%%%%%%%%%%%%%%%%%%%%%%%%%%%%%%%%%%%%%%%%%%%%%%%%%%%%%%%%%%%%%%%%%%%%%%%%%%%%%%%%%%%%%%
%
\section{Introduction}

Precision measurements of the cosmic microwave background radiation (CMB) have established inflation as the predominant framework for explaining the observed flatness and homogeneity of the Universe \cite{Starobinsky:1980te,Guth:1980zm,Linde:1981mu,Albrecht:1982wi}. Inflation also provides a natural mechanism to produce primordial perturbations which seed the observed large scale structures. The two main observables from CMB are the tensor-to-scalar ratio $r$ and the spectral index $n_s$ which can be exploited to constrain the landscape of viable inflationary scenarios.  

The latest data release from the Atacama Cosmology Telescope (ACT) updated the constraints on inflationary cosmology \cite{ACT:2025fju,ACT:2025tim}. When combined with the Year 1 data from DESI \cite{DESI:2024mwx,DESI:2024uvr}, the constraint on the spectral index is $n_s = 0.9743 \pm 0.0034$, a $2\sigma$ shift compared to the previous Planck 2018 dataset $n_s = 0.9651 \pm 0.0044$ \cite{Planck:2018jri}. This implies that, for the first time, the original Starobinsky model is disfavored at $2\sigma$ confidence level. The new results from ACT already attracted attention from the community and new proposals have been recently made \cite{Kallosh:2025rni,Aoki:2025wld,Dioguardi:2025vci, Salvio:2025izr,Gialamas:2025kef}.

Linear inflation predicts $n_s = 1-\frac{3}{2 N}$. Hence, for $N = 60$ $e$-folds of inflation, $n_s =0.975$ which lays remarkably close to the observed Planck+ACT value. The attractor behavior of linear inflation \cite{Racioppi:2018zoy} makes the model appealing for understanding the dynamics of early universe acceleration. However, the prediction for the tensor-to scalar ratio  $r = 0.0667$ lays outside the $2\sigma$ experimental bounds and is strongly disfavored by current observations.
The model needs then to be revisited and embedded in a different framework if we want to restore its viability.\\
In this letter we show that the  Palatini formulation of gravity  represents one way to do it.
In the Palatini formalism the connection is a priori taken to be independent from the metric. If the gravity sector is non-minimally coupled to matter fields then phenomenological predictions will be different from predictions of metric gravity \cite{Koivisto:2005yc,Bauer:2008zj}.
At the same time, in the presence of a quadratic term $R^2$ in the action, Palatini gravity has the appealing feature to arbitrarily lower the tensor-to-scalar ratio $r$, without affecting the other inflationary observables \cite{Enckell:2018hmo}. Higher-order and extended Palatini models have also been studied and have proven to have appealing features to generate viable inflation \cite{Dioguardi:2021fmr,Dioguardi:2022oqu, Dioguardi:2023jwa,Dimopoulos:2025fuq,Bostan:2025vkt}.
In this letter we specifically consider (i) chaotic inflation in presence of a non-minimal coupling $\xi$ between the inflaton and the Ricci scalar (ii) minimally coupled chaotic inflation in presence of an higher-order $\alpha R^2$ term. \\
The letter is organized as follows: in section \ref{sec:Palatini} we introduce the general action and briefly describe how to compute inflationary observables. In section \ref{sec:nonminimal_inflation} we study the non-minimally coupled chaotic models and prove that a quadratic Jordan frame potential with a linear non-minimal coupling is favored by the current ACT dataset. In section \ref{sec:R^2} we consider minimally coupled chaotic models in presence of a $\alpha R^2$ term in the action and show that linear inflation is currently the favored model. Finally in \ref{sec:conclusions} we draw our conclusions.

%%%%%%%%%%%%%%%%%%%%%%%%%%%%%%%%%%%%%%%%%%%%%%%%%%%%%%%%%%%%%%%%%%%%%%%%%%%%%%%%%%%%%%%%%%%%%%%

\section{Palatini Inflation}\label{sec:Palatini}

We begin by considering an action of the form~\cite{Enckell:2018hmo, Antoniadis:2018ywb, Gialamas:2023flv}
\be \label{action1}
  S = \int\dd^4 x \sqrt{-g} \left[ \frac{1}{2} F\left( \phi, R \right) - \frac{1}{2} g^{\mu\nu} \partial_\mu \phi \partial_\nu \phi - V(\phi) \right] \,,
\ee
where $g$ is the determinant of the spacetime metric $g_{\mu\nu}$ and $V(\phi)$ is the Jordan frame potential of the inflaton $\phi$. We have introduced the function
\be \label{action1}
  F\left( \phi, R \right) = A(\phi) R + \alpha R^2 \,,
\ee
where $\alpha$ is a positive constant and $R=g^{\mu\nu} R^{\rho}_{\ \mu\rho\nu}(\Gamma, \partial\Gamma)$ is the curvature Ricci scalar. We consider a general non-minimal coupling function $A(\phi)$ between the inflaton and gravity in the action and a quadratic curvature term $\alpha R^2$, while assuming a canonical kinetic term for the inflaton field. Throughout this work, we adopt the Palatini formulation of gravity, in which the metric $g_{\mu\nu}$ and the connection $\Gamma$ are treated as independent variables. Additionally, we impose the condition that the connection is torsion-free, i.e., $\Gamma^\lambda_{\mu\nu} = \Gamma^\lambda_{\nu\mu}$.

In order to make contact with inflationary observables, it is easier to work in the Einstein frame. After eliminating the $R^2$ term with the help of an auxiliary field $\chi \equiv 2 \alpha R$, performing a Weyl transformation $g_{\mu\nu} \rightarrow \Omega^2 g_{\mu\nu} = [ \chi + A(\phi) ] g_{\mu\nu}$  and utilizing a field redefinition of the form
\be 
    \label{eq:field_redef}
  \left(\frac{\dd \phi}{\dd \varphi}\right)^2 = A(\phi) ( 1 + 8 \alpha \bar{U}(\phi) ) \,, \qquad \bar{U}(\phi) \equiv \frac{V(\phi)}{[A(\phi)]^2}\,,
\ee
one finds~\cite{Enckell:2018hmo, Antoniadis:2018ywb}
\be 
  S = \int\dd^4 x \sqrt{-g} \left[ \frac{1}{2} R - \frac{1}{2} \left( \partial \varphi \right)^2 - U(\varphi)  \right] \,, 
\ee
where
\be 
    U \equiv \frac{\bar{U}}{1 + 8 \alpha \bar{U}}\,,
\ee
and we have neglected a higher-order kinetic term of the form $\frac{\alpha}{2} \left( 1 + 8 \alpha \bar{U}(\varphi) \right) \left( \partial \varphi \right)^4$, which has been shown not to play a significant role during inflation~\cite{Tenkanen:2020cvw, Karam:2021sno}. It is evident that, irrespective of the form of $\bar{U}$, the $\alpha R^2$ term decreases the height of the potential in the Einstein frame. At large field values, the effective potential flattens and asymptotically approaches the value $1/(8\alpha)$.

Regarding the inflationary observables, the leading order expressions for the amplitude of the scalar power spectrum $A_s$ and scalar spectral index $n_s$ as functions of $\phi$ do not explicitly explicitly on $\alpha$~\cite{Enckell:2018hmo}
\be \label{eq:As-ns}
24 \pi^2 A_s = \frac{U}{\epsilon_U} = \frac{\bar{U}}{\epsilon_{\bar{U}}} \,, \qquad n_s = 1 - 6 \epsilon_U + 2 \eta_U = 1 - 6 \epsilon_{\bar{U}} + 2 \eta_{\bar{U}}\,,
\ee
with the potential slow-roll parameters defined as
\be 
\epsilon_U = \frac{1}{2} \left( \frac{U'}{U} \right)^2 \,, \qquad \epsilon_{\bar{U}} = \frac{1}{2} \left( \frac{\bar{U}'}{\bar{U}} \right)^2 = \epsilon_U \vert_{\alpha = 0}\,,
\ee
\be 
\eta_U =  \frac{U''}{U} \,, \qquad \eta_{\bar{U}} =  \frac{\bar{U}''}{\bar{U}}  = \eta_U \vert_{\alpha = 0}\,.
\ee
Note that the measured value of the amplitude is $A_s = 2.1 \times 10^{-9}$~\cite{Planck:2018jri}.
On the other hand, the tensor power spectrum explicitly depends on $\alpha$
\be 
A_T = \frac{2}{3 \pi^2} U = \frac{2}{3\pi^2} \frac{\bar{U}}{1 + 8 \alpha \bar{U}}\,.
\ee
As a result, the tensor-to-scalar ratio becomes 
\be 
r = 16 \epsilon_U = \frac{\bar{r}}{1 + 8 \alpha \bar{U}}  = \frac{\bar{r}}{ 1+ 12 \pi^2 A_s \bar{r} \alpha } \,, \label{eq:r}
\ee
where in the last equality we used eq.~\eqref{eq:As-ns} and $\bar{r}=16\epsilon_{\bar{U}}$ is the tensor-to-scalar ratio of the same model but without the $\alpha R^2$ term. 
We thus see that increasing the value of $\alpha$ allows the value of $r$ to be suppressed within a given inflationary model, without altering the prediction for $n_s$.
In the following, we first consider the case of a non-minimal coupling $A(\phi) \neq 1$ in the limit where the $R^2$ term is absent ($\alpha = 0$) and then we focus on the case of a minimally coupled inflaton ($A(\phi) = 1$) with $\alpha \neq 0$.

\subsection{Non-minimally coupled models} %of inflation in Palatini gravity} 
\label{sec:nonminimal_inflation}

\begin{figure*}[t!]
\begin{center}\label{fig:slow-roll}
\includegraphics[width=1\textwidth]{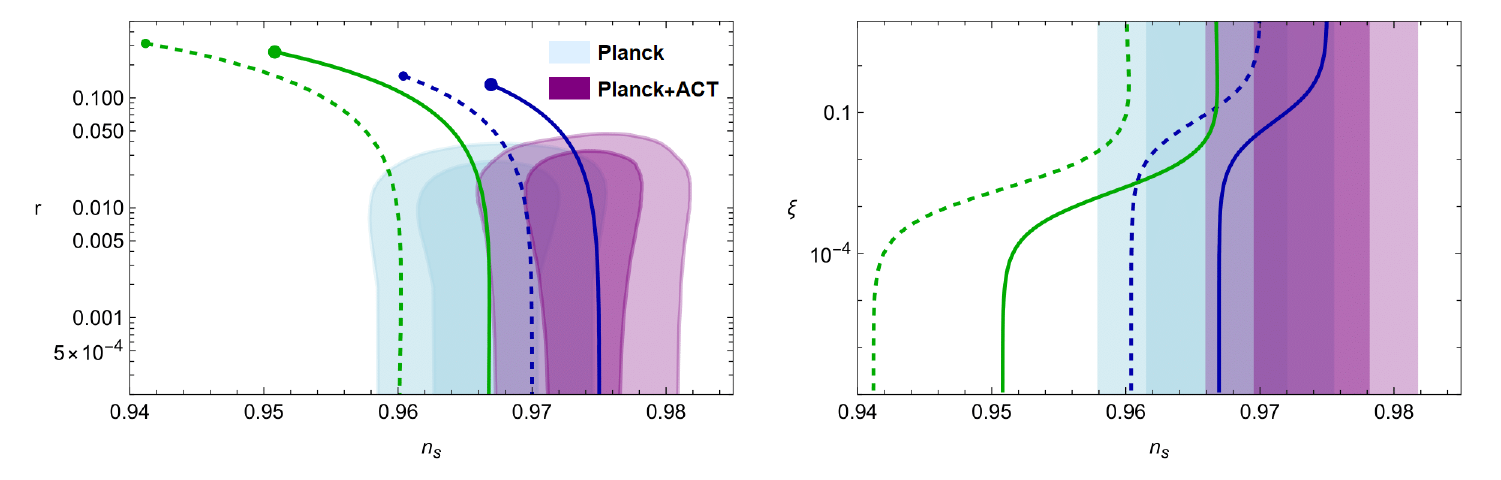}
\caption{\textbf{Left}: $r$ vs. $ns$ for the chaotic model defined by eqs.\eqref{eq:f}-\eqref{eq:V}, for $n=1$ (blue) and $n=2$ (green), with $N=50$ (dashed) and $N = 60$ (thick). For reference, we show the predictions for $\xi=0$ (i.e.~the simple quadratic and quartic potential) with dots in the same color code. 
\textbf{Right:} $\xi$ vs. $n_s$ for  for $n=1$ (blue) and $n=2$ (green), with $N=50$ (dashed) and $N = 60$ (thick).
The green and blue lines are obtained by varying $10^{-7}\lesssim\xi\lesssim 10$. The amplitude of the curvature power spectrum is fixed to the observed value~\cite{Planck:2018jri}. The purple areas present the 1, 2$\sigma$ constraints from the Planck+ACT data \cite{ACT:2025tim}, while the light blue ones represent the older constraints from Planck \cite{BICEP:2021xfz}. 
}
\label{fig:U:vs:zetabar}
\end{center}
\end{figure*}

In this section, we consider the subclass of models for which $\alpha=0$. The Einstein frame potential is given by $U = \Bar{U}$ defined in eq.~\eqref{eq:field_redef}, and the relation between the Jordan frame scalar field $\phi$ and the canonical Einstein frame field $\varphi$ reduces to
\be
\label{eq:NM_field_redef}
\qty(\frac{\dd\phi}{\dd\varphi})^2 = A(\phi)\,.
\ee
Next, let us consider a general class of chaotic non-minimally coupled models
\begin{eqnarray}
\label{eq:f} 
A(\phi) &=& 1 + \xi \phi^n \, ,\\
F(\phi,R) &=& \left( 1 + \xi \phi^n \right) R \,, \\
V(\phi) &=& \lambda_{2n} \phi^{2n} \label{eq:V}
\end{eqnarray}
where $n>0$ and $\lambda_{2n}$ is a dimensionless coupling. The constraint on the amplitude of scalar perturbations~\cite{Planck:2018jri} allows us to fix the value of $\lambda_{2n}$.
After integration, the scalar field redefinition~\eqref{eq:NM_field_redef} yields
\begin{equation}
\label{eq:hypergeom}
\varphi = \phi F \left(\tfrac{1}{2}, \tfrac{1}{n}; \tfrac{n+1}{n},-\xi \phi^n \right) \,,
\end{equation}
where $F$ is a hypergeometric function. For general $n$ it is difficult to obtain an analytic expression for the potential $U(\phi(\chi))$, but for e.g.\ $n=2$ we recover~\cite{Bauer:2008zj}
\begin{equation}
U(\varphi) \simeq \frac{\lambda_4}{\xi^2} \left(1-8 e^{-2 \sqrt{\xi} \varphi} \right) \,,
\label{eq:Palatini:attractor:4}
\end{equation}
while $n=1$ yields
\begin{equation}
U(\varphi) \simeq \frac{\lambda_2}{\xi^2} \left(1- \frac{8}{(2 + \xi \varphi)^2}\right) \,.
\end{equation}
For $\xi\to\infty$, we find that
\bea\label{eq:chaotic_ns}
n_s &\simeq& 1 - \left( 1+\frac{n}{2} \right) \frac{1}{N} \,, \\ \nonumber
r &\simeq& 0 \, ,
\eea

%%%%%%%%%%%%%%%%%%%%%%%%%%%%%%%%%%%%%%%%%%%%%%%%%%%%%%%%%%%%%%%%%%%%%%%%%%%%%%%%%%%%%%%%%%

As we can see from~\eqref{eq:chaotic_ns}, taking into account the Planck+ACT dataset prediction $n_s = 0.9743 \pm 0.0034$, the best-fitting model corresponds to the case $n=1$, which for $N=60$ gives $n_s = 0.975$. We plot the numerical analysis for the slow-roll results for $n=1$ (blue) and $n=2$ (green) with $N=50$ and $N=60$ (dashed) in Fig.1. The green and blue lines are obtained by varying $10^{-7}\lesssim\xi\lesssim 10$. The amplitude of the scalar curvature power spectrum is fixed to the observed value $A_s\sim 2.1\cdot 10^{-9}$. The purple areas represent the 1, 2$\sigma$ constraints from the Planck+ACT data \cite{ACT:2025tim}, while the light blue ones represent the older constraints from Planck \cite{BICEP:2021xfz}.  
We see that the Planck+ACT data disfavors the $n=2$ model at a confidence level $\gsim 2\sigma$ while it strongly favors the $n=1$ model at couplings $\xi \gtrsim 0.1$ for $50<N<60$.

%%%%%%%%%%%%%%%%%%%%%%%%%%%%%%%%%%%%%%%%%%%%%%%%%%%%%%%%

\subsection{$R^2$ models}\label{sec:R^2}

\begin{figure*}[t!]
\begin{center}\label{fig:slow-roll}
\includegraphics[width=1\textwidth]{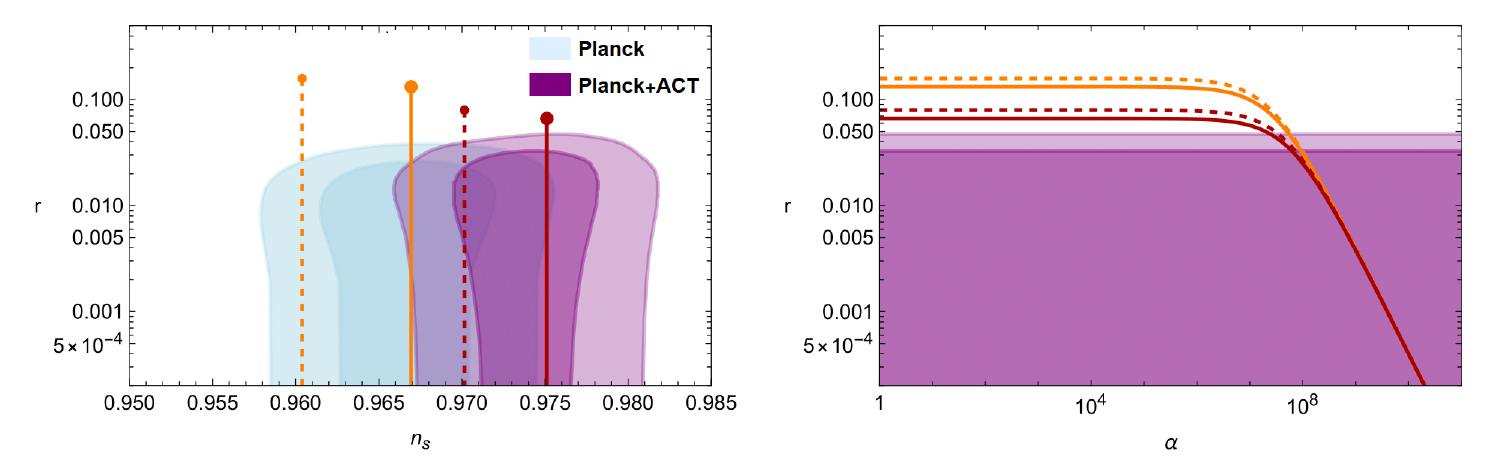}
\caption{\textbf{Left}: $r$ vs. $ns$ for the quadratic (orange) and linear (red) Jordan frame potential with $N_e=50$ (dashed) and $N_e = 60$ (thick). 
\textbf{Right:} $r$ vs. $\alpha$ for the quadratic (orange) and linear (red) Jordan frame potential with $N_e=50$ (dashed) and $N_e = 60$ (thick). The orange and red lines are obtained by varying $0 < \alpha\lesssim10^{10}$. The amplitude of the scalar curvature power spectrum is fixed to the observed value $A_s \sim 2.1\cdot 10^{-9}$. The purple areas present the 1, 2$\sigma$ constraints from the Planck+ACT data \cite{ACT:2025tim}, while the light blue ones represent the older constraints from Planck \cite{BICEP:2021xfz}.
}
\label{fig:R2}
\end{center}
\end{figure*}

We now turn our attention to the case where the scalar field is minimally coupled but the gravitational sector contains a term quadratic in curvature. As before, we assume simple chaotic potentials. The model functions are:
\bea
A(\phi) &=& 1 \,, \\  
F(\phi,R) &=& \alpha R^2 \,, \\
V(\phi) &=& \lambda_{2n} \phi^{2n} \,.
\eea
This time, the field redefinition gives
\be 
    \varphi = \phi F \left(\tfrac{1}{2}, \tfrac{1}{2n}; 1 + \tfrac{1}{2n},-8 \alpha \lambda_{2n} \phi^{2n} \right) \,.
\ee
By choosing $n$, one can solve the above equation in terms of $\phi$ and express the Einstein frame potential $U(\varphi)$ in terms of the canonical field $\varphi$. For example, for $n=1$, one finds~\cite{Antoniadis:2018ywb, Gialamas:2020snr, Karam:2021sno}
\be 
U(\varphi) = \frac{\tanh^2\left( 4 \sqrt{\alpha} \lambda_2 \varphi \right)}{8 \alpha} \,.
\ee
Next, we compute the inflationary observables and express them in terms of the number of $e$-folds. As mentioned above, the $\alpha R^2$ term in the action allows to arbitrarily lower the tensor-to-scalar ratio, while leaving all the other observables unaffected. For $n=1/2$ we find
        \bea 
            r &=& \frac{4}{N + 8 \sqrt{2}\, \alpha\, \lambda_1 N^{3/2}} \,, \\
            n_s &=& 1 - \frac{3}{2 N} \,, \\
            A_s &=& \frac{\lambda_1 N^{3/2}}{3 \sqrt{2}\, \pi^2} \,,
        \eea
while for $n=1$ we have
        \bea 
            r &=& \frac{8}{N_e + 8 \, \alpha \, \lambda_2 N^2} \,, \\
            n_s &=& 1 - \frac{2}{N} \,, \\
            A_s &=& \frac{\lambda_2 N^2}{3 \pi^2} \,.
        \eea
For $\alpha \rightarrow 0$, we recover the linear inflation limit in the case of $n=1/2$ and the quadratic inflation limit in the case of $n=1$. On the other hand, for $\alpha \rightarrow \infty$, we have $r \rightarrow 0$, and we can thus bring both linear and quadratic inflation predictions for $r$ in agreement with the constraint $r < 0.036$~\cite{Planck:2018jri, BICEP:2021xfz}.

We plot in Fig.~\ref{fig:R2} the results for the quadratic (orange) and linear (red) Jordan frame potential with $N_e=50$ (dashed) and $N_e = 60$ (thick). 
The orange and red lines are obtained by varying $0 < \alpha\lesssim10^{10}$. The amplitude of the scalar curvature power spectrum is fixed to the observed value $A_s \sim 2.1\cdot 10^{-9}$. The purple areas represent the 1, 2$\sigma$ constraints from the Planck+ACT data~\cite{ACT:2025tim}, while the light blue ones represent the older constraints from Planck~\cite{BICEP:2021xfz}. We notice that with the Planck+ACT data, quadratic inflation is disfavored at a confidence level $\gsim 2\sigma$ while linear inflation is strongly favored in the presence of an $\alpha R^2$ term in the action for $\alpha \gtrsim 10^8$ for $50<N<60$.

\section{Conclusions}\label{sec:conclusions}
 
The latest ACT results indicate a preference for a scalar spectral index $n_s \simeq 0.9743$, remarkably close to the prediction of linear inflation, $n_s = 1 - \frac{3}{2N}$, which yields $n_s = 0.975$ for $N = 60$ $e$-folds. However, the corresponding prediction for the tensor-to-scalar ratio, $r \simeq 0.0667$, lies above current observational bounds.

In this work, we explored how such tension can be resolved within the framework of Palatini gravity, considering two classes of models based on chaotic potentials. In the first class, we introduced a non-minimal coupling $\xi$ between the inflaton and gravity. We showed that a quadratic potential with a linear non-minimal coupling exhibits the linear inflation attractor behavior, with the value of $r$ decreasing as $\xi$ increases, while $n_s$ lies within the new constraints for $\xi \gtrsim 0.1$. In the second class, we supplemented chaotic potentials with an $\alpha R^2$ term. In this case, we found that a linear potential can yield a suppressed tensor-to-scalar ratio consistent with current constraints, provided $\alpha \gtrsim 10^8$.

These results demonstrate that Palatini gravity offers a natural mechanism to reconcile linear inflation–like predictions for $n_s$ with the observed upper bounds on $r$, without requiring significant departures from simple potential forms.

\section*{Acknowledgements}
This work was supported by the Estonian Research Council grants PRG1055, PSG761, RVTT3, RVTT7 and the CoE program TK202 ``Foundations of the Universe'’. We thank I.D. Gialamas and A. Racioppi for useful discussions.

\bibliography{references}

\end{document}